# RAPID-CYCLING SYNCHROTRON WITH VARIABLE MOMENTUM COMPACTION*

Y. Alexahin#, FNAL, Batavia, IL 60510 U.S.A.
D. J. Summers, University of Mississipi-Oxford, University, MS 38677 U.S.A.

*Abstract*

There are conflicting requirements on the value of the momentum compaction factor during energy ramping in a synchrotron: at low energies it should be positive and sufficiently large to make the slippage factor small so that it is possible to work closer to the RF voltage crest and ensure sufficient RF bucket area, whereas at higher energies it should be small or negative to avoid transition crossing. In the present report we propose a lattice with a variable momentum compaction factor and consider the possibility of using it in a high repetition rate proton driver for a muon collider and neutrino factory.

## INTRODUCTION

High power proton synchrotrons must avoid transition energy crossing to reduce beam losses and high background radiation levels. The most radical solution is the implementation of a lattice with negative momentum compaction factor, $\alpha_p < 0$, so that the transition energy is imaginary.

A possible drawback of this solution is correspondingly large values of the slippage factor $\eta$ at the earlier part of the energy ramp. Here we define $\eta$ as:

$$\eta = \frac{1}{\gamma^2} - \alpha_p \qquad (1)$$

The RF bucket area decreases with $\eta$ as $1/\eta^{1/2}$, so a larger RF voltage is required with $\alpha_p < 0$, especially when the energy gain per turn reaches its regular value but the beam longitudinal emittance is still large. A possible solution to this problem is $\alpha_p$ variation during the ramp as discussed in the following Section.

As an example we consider a Rapid Cycling Synchrotron (RCS) with variable momentum compaction (VMC) for Fermilab Project X [1]. The ultimate goal of this project is to provide 2 MW proton beam from the Main Injector (MI) at 120 GeV for neutrino experiments and up to 4 MW of proton beam power at 8 to 20 GeV for a future Neutrino Factory and Muon Collider [2].

The first stage of the project involves a 2.6 to 3 GeV proton linac. We take the lower bound - 2.6 GeV - for the injection energy in our RCS. Following D. McGinnis [3], we consider the possibility of injecting beam into the MI above its transition energy of 20.4 GeV. Thus the RCS final energy should be at least 21 GeV.

To achieve the more distant Project X goal – a 4 MW

___________________________________________
* Work supported by Fermi Research Alliance, LLC under Contract DE-AC02-07CH11359 with the U.S. DOE and NSF Grant 757938.
#alexahin@fnal.gov

Muon Collider proton driver – 200 μA or 1.25x10^15 of 21 GeV protons per second will be required. With a realistic number of ~2·10^13 ppp the RCS repetition rate has to be as high as 60 Hz.

The last Section of this Report presents some ideas on the design of fast ramping magnets and their driving circuits which may help to achieve such a high repetition rate.

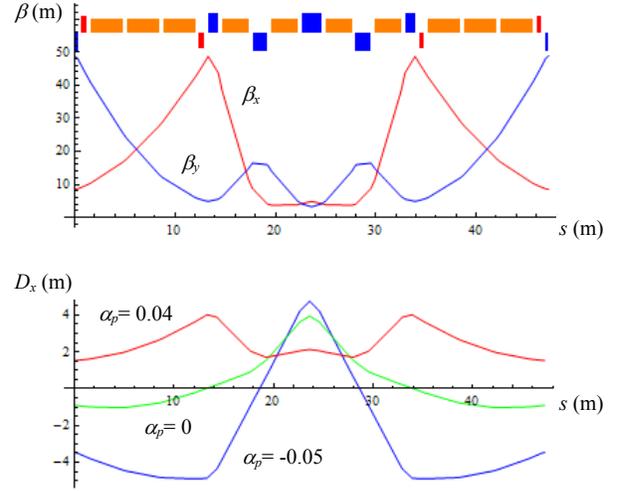

Figure 1: Optics functions in the VMC cell: $\beta$-functions at $\alpha_p = 0$ (top); the dispersion function at indicated values of $\alpha_p$ (bottom).

## VMC LATTICE

With $\alpha_p < 0$, the synchrotron tune initially can be very high, so it is advantageous to place RF cavities in dispersion-free sections, thus eliminating systematic synchro-betatron coupling. Since dispersion suppressors produce a positive contribution to $\alpha_p$, their number should be minimized to allow negative momentum compaction in the whole lattice. Therefore, we choose a racetrack configuration for our RCS.

### Arc Cell

The arc cell design (see Fig. 1) follows that proposed for a muon collider [4]. It resembles the PS2 design [5] but has some important differences: there is no free spaces (except those required for correctors) so the dipole packing is potentially higher; the maximum dispersion is reached at a location where both $\beta$-functions are small facilitating an independent control of $\alpha_p$ and betatron tunes.

The number of cells per arc $N_{cell}$ is a trade-off between the desire to have as large an $\alpha_p$ range as possible which calls for a high dispersion (also beneficial for space-charge tuneshift reduction) and limitations on the aperture and circumference. Assuming the horizontal aperture half-width 35 mm and the maximum momentum deviation in the beam $\delta_p \sim 0.6\%$ we may allow for dispersion values of up to 5 m. Limiting the bending field to 1.7 T and maximum quad gradient by 18 T/m (and inserting 0.4 m spaces between all elements) we find the optimum cell length $L_{cell}$ = 47.2 m and $N_{cell}$ = 4.

To achieve intrinsic cancellation of the 3$^{rd}$ order resonances excited by chromaticity correction sextupoles (shown at the top of Fig. 1 with red) the phase advance per cell $\mu$ = 270° was chosen for both planes.

As already noted it is easy with this lattice to vary $\alpha_p$ while keeping the phase advances fixed. With $\alpha_p$ changing from -0.05 to 0.04 there is little change in the $\beta$-functions and all quad gradients remain within specified limits. However, the dispersion function varies significantly (Fig. 1 bottom) crossing zero at sextupole locations at some values of $\alpha_p$. Additional study will show if extra sextupoles are necessary.

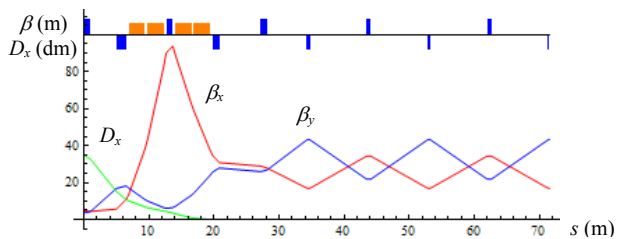

Figure 2: Optics functions in the dispersion suppressor and a half of straight section for $\alpha_p$ = -0.002

*Dispersion Suppressor & Straight Section*

A simple solution for the dispersion suppressor optics is shown in Fig. 2. A larger value of $\beta_x$ at the bends was chosen deliberately to reduce the required bending angle. This does not require more aperture than in the arcs since $D_x$ is already small at these bends. With dispersion suppressors added the bending angles and the dispersion function in the arcs are reduced by some 15% compared to that shown in Fig. 1.

The straight section length is determined by the RF system requirements. With the asymmetric rise/fall cycle and the double frequency magnet power harmonic described in the next Section, the maximum energy gain is ~5 MeV/turn in the 60 Hz case. Assuming the RF cavity design similar to that used in the Fermilab Booster and noting that now the lower frequency bound is 51 MHz instead of 30 MHz, we estimate the average accelerating gradient to be ~ 38 kV/m so that the total RF system length is ~ 160 m. There are proposals which promise even higher gradients [6].

We set the RCS circumference at 663.835 m or 1/5 of the MI circumference which gives us 188 m free space in the straight sections for RF, injection and extraction kickers, and dampers.

Due to a large contribution from the straights to the circumference the available $\alpha_p$ range becomes smaller: from -0.02 to +0.018. Still, variation of $\alpha_p$ during the ramp from negative at injection and bunching to positive and then to again negative allows one to reduce the required RF voltage. If $\alpha_p$ can be controlled with absolute precision of better than 0.005 this reduction can be as large as 20% resulting in an RF voltage of 6 MV.

The straight section quadrupoles can be used to set the desired values of betatron tunes. The optics shown in Fig.2 correspond to $Q_x$= 8.44, $Q_y$= 8.46. The natural chromaticity of the lattice is $Q_x'$ = -17.4, $Q_y'$ = -12.0. It is worthwhile to note that no change is necessary in normalised strength of magnets in the dispersion suppressor and straight section with varying $\alpha_p$. However, with different from the nominal value $\alpha_p$ the dispersion function exhibits significant beat in the arcs.

## RCS MAGNET SYSTEM

A 60 Hz ring with 1.7 T dipoles is challenging. The ISIS ring at Rutherford Lab delivers an 800 MeV proton beam of 2.5x10$^{13}$ protons per pulse at 50 Hz, but the dipoles only run up to 0.7 T [7]. A 1250 Hz pulsed wiggler at Brookhaven Laboratory has achieved 2.1 T using a vanadium permendur lamination yoke and a 4mm gap [8]. Amperage and voltage of dipole magnets are as follows [9]:

$$I = B h / \mu^0 N, \quad V = \pi B f N w \ell, \quad (2)$$

where $h$, $w$, and $\ell$ are the gap height, width, and length, $N$ is the number of coil turns, and $f$ is frequency. Using an offset White circuit [10] rather than an LC circuit halves the voltage. With a 0.05m x 0.1m x 3m magnet gap, 1.7 T B field, 120 Hz fall frequency, and 18 coil turns, one gets 3800 amps and 3500 volts, which is large but probably doable. The voltage to ground can be halved to 1750 via a center tap.

To give more time for RF acceleration, an asymmetric cycle is used with a 40 Hz rise and a 120 Hz fall. A double frequency harmonic is added to the power supply so that the RF cavities can deliver their maximum gradient during a larger fraction of the rise cycle.

Small copper wire around a stainless steel cooling tube with flowing water is used to reduce eddy currents in the magnet conductor. The new ISIS choke at Rutherford Lab uses this method [11]. The cable is made by Trench Ltd. The Japan Accelerator Research Complex uses thin aluminium wire around a stainless steel cooling tube [12].

To minimize stored energy ($B^2/2\mu$) and hysteresis and eddy current losses in the dipole yokes, thin 3% grain oriented silicon steel laminations are used [13]. At 1.7 T in the grain direction $\mu/\mu^0$ has a value of 30000, but the value falls to 3000 at 10° away from the grain direction. At 1.3 T in the grain direction $\mu/\mu^0$ has a value of 47000, but the relative permeability falls to 14000, 540, and 2300 at 10°, 55°, and 90° away from the grain direction, respectively [14]. The 55° minimum comes from the long diagonal (111) of the steel crystal. At 1.7 T, $\mu/\mu^0$ for 0.0025% ultra low carbon steel is 600 [15] or 50 times less than high grade grain oriented silicon steel. If stored energy density ($B^2/2\mu$) in the dipole yoke is not minimized, power supply voltages and amperages will rise. To take advantage of grain oriented steel, large 3-phase transformers are built with 45° mitred joints [16]. So magnet laminations may need to be laid out as shown in Fig. 3. Tests of fast ramping prototype dipole magnets with 3% grain oriented silicon steel laminations are in progress at the University of Mississipi-Oxford.

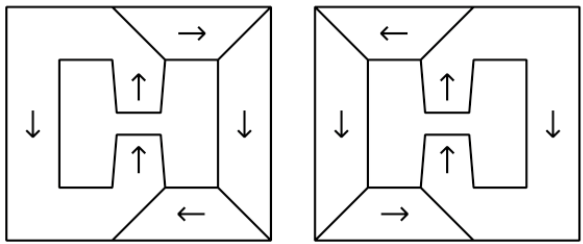

Figure 3: Alternating dipole laminations of 3% grain oriented silicon steel. The arrows show the **B** field direction and the grain direction. The **B** field and the grain directions are parallel.

*Power Losses in Magnets*

The mass of the magnets is about 600 metric tons and the steel laminations consume 0.9 watts/kg with a 1.5 T average field and 60 Hz due to eddy currents and hysteresis losses. This works out to a total of 540 kilowatts for all the laminations. $I^2R$ losses for 18 turns of 3cm x 3cm stranded copper cables carrying 3800 amps of sinusoidal current for 270 meters of magnets works out to 1300 kilowatts.

The power dissipated in a round wire by eddy currents per unit volume is $(2\pi f B d)^2 / 24\rho$, where $f$ is frequency, $B$ is magnetic field, $d$ is diameter, and $\rho$ is resistivity [13]. So a copper volume of 9 m$^3$, 0.002m diameter copper wire, an 0.2 T fringe field, and 60 Hz results in eddy current losses of 500 kilowatts. Total losses are approximately 2.3 megawatts for all magnets.

## SUMMARY

The basic parameters of the proposed machine are summarized in Table 1.

The presented design is a preliminary one and leaves many issues untouched, such as suppression of coherent beam instabilities at the considered intensity. We plan to address these issues in the future.

Table 1: RCS Parameters

| | | |
|---|---|---|
| Energy | GeV | 2.6 → 21 |
| Circumference | m | 663.835 |
| Maximum B field | T | 1.7 |
| Repetition rate | Hz | 60 |
| RF frequency | MHz | 50.9 → 52.8 |
| Maximum RF voltage | MV | 6 |
| RF bucket area | eV·s | 0.6 |
| Protons/bunch | 10$^{11}$ | 2 |
| Bunch number | - | 112 |
| 95% Emittance | π·mm·mrad | 20 |
| Space charge tuneshift (Gaussian distribution) | - | 0.2 |
| Beam power | MW | 4.5 |